\documentclass[referee]{raa}   
\usepackage[utf8]{inputenc}
\DeclareUnicodeCharacter{FF0C}{,}
\renewcommand{\figurename}{Figure} 
\usepackage{graphicx,times}             
\usepackage{natbib}
\usepackage{amssymb,amsmath}
\bibpunct{(}{)}{;}{a}{}{,}

\usepackage[pagebackref=true]{hyperref}
\usepackage{orcidlink}
\usepackage{xcolor}

\begin{document}

  \title{Unveiling Galactic substructures with M Giant stars: A kinematic and chemical study based on LAMOST DR9, Gaia DR3 and APOGEE DR17}

   \volnopage{Vol.0 (20xx) No.2022-0433}      
   \setcounter{page}{1}          

   \author{Longfei Ding
      \inst{1} \orcidlink{0009-0007-6842-8117}
   \and Jing Li
      \inst{1} \orcidlink{0000-0002-4953-1545}
   \and Xiang-Xiang Xue 
      \inst{2,3} \orcidlink{0000-0002-0642-5689}
    \and Hao Tian
          \inst{2,3} \orcidlink{0000-0003-3347-7596}
    \and Zhengzhou Yan
      \inst{1} \orcidlink{0000-0003-3571-6060}
    \and Gang Zhao
      \inst{2, 5} \orcidlink{0000-0002-8980-945X}
   }

   \institute{School of Physics and Astronomy, China West Normal University, Nanchong 637002, China; {\it lijing@shao.ac.cn, zzyan@bao.ac.cn}\\
        \and
             National Astronomical Observatories, Chinese Academy of Sciences, Beijing 100101, China;{\it xuexx@nao.cas.cn}\\
        \and
            Institute for Frontiers in Astronomy and Astrophysics, Beijing Normal University, Beijing 102206, China\\
        \and
            School of Astronomy and Space Science, University of Chinese Academy of Sciences, Beijing 100049, China;\\
\vs\no
   {\small Received 20XX Month Day; accepted 20XX Month Day}}

\abstract{Based on the updated M giant star catalog selected from LAMOST DR9, we identify substructures within the integrals-of-motion space through Friends-of-Friends clustering algorithm. We obtain members belonging to several known substructures: the Sagittarius stream, Galactic Anticenter Substructure (GASS), Gaia-Enceladus-Sausage (GES), Splash, and the high-$\alpha$ disk. Furthermore, we also identify two groups which cannot be clearly associated with previously known substructures. Our findings confirm the existence of metal-rich constituents within the GES, representing newly formed stars that originated from the metal-enriched gas delivered during the GES merger event and subsequently evolved. Additionally, this study further expands the sample of GASS, high-$\alpha$ disk, and Splash stars. Analysis of these metal-rich M giant stars as members of the GES, Splash, and high-$\alpha$ disk components supports an evolution scenario for the early Milky Way, as proposed by previous studies. In this scenario, stars initially formed in a high-$\alpha$ primordial disk were dynamically heated by the massive accretion event (GES). This process redistributed stellar orbits, creating the Splash population, while the undisturbed portion of the primordial disk persisted as the present-day high-$\alpha$ disk component.
}

   \authorrunning{Longfei Ding et al.}            
   \titlerunning{Unveiling Galactic substructures with M Giants}  
   \maketitle
%
%
\section{Introduction}\label{sect:intro}
Galactic archaeology plays a crucial role in uncovering the formation and evolutionary history of the Milky Way \citep{Helmi_2020}. According to the $\Lambda$CDM model, the Galaxy primarily formed through hierarchical mass assembly, where satellite galaxies merged with the Milky Way \citep{Horta_2023}.  These events created substructures in the Galactic halo and disk, characterized by stars with consistent kinematic, dynamical, and chemical properties \citep{Helmi_2020}. Studying these substructures helps reconstruct the Galaxy’s merger history \citep{Helmi_2018, Massari_2019, Yang_2019a, Myeong_2019, Koppelman_2019b, Bonaca_2020, Li_2021, Zhao_2021, Wang_2022}.

Utilizing data from photometric and spectroscopic surveys, such as the Sloan Digital Sky Survey \citep[SDSS;][]{Tork_2000, Ahumada_2020}, the Two Micron All Sky Survey \citep[2MASS;][]{Skrutskie_2006}, Pan-STARRS1 \citep[PS1;][]{Chambers_2016}, the Large Sky Area Multi-Object Fiber Spectroscopic Telescope \citep[LAMOST;][]{Zhao_2012, Yan_2022}, and Gaia \citep{Gaia_2023}, more and more substructures have been identified in the Milky Way \citep{Newberg_2016, Belokurov_2018, Haywood_2018, Myeong_2018, Ibata_2021, Mateu_2023}. Recent studies have used diverse datasets and methods to identify Galactic substructures. For example, using various types of stellar samples such as K giants, BHB stars, and RR Lyrae stars, many substructures within the Galactic halo, including the Sagittarius (Sgr) stream, Gaia-Enceladus-Sausage (GES), Sequoia, Helmi stream, Thamnos and LMS-1, have been identified \citep{Yang_2019b, Yuan_2020, Naidu_2020, Wang_2022, Sun_2024}. Meanwhile, applying various methods (such as Friends-of-Friends (FoF), HDBSCAN, StarGO and STREAMFINDER) to sample in phase space\footnote{The three-dimensional position and three-dimensional velocity space.} and integrals of motion (IoM) space\footnote{$E$ is the total energy of the star,  $L_z$ is the component of the star's orbital angular momentum in the vertical direction, and ($J_R$, $J_\Phi$, $J_z$) are three actions in an axisymmetric system. All these parameters are integrals of motion \citep{Ollongren_1965, Binney_2008, Binney_2012, Helmi_2020}.} based on position-velocity information has made the identification of Galactic substructures more accurate and efficient \citep{Belokurov_2020, Necib_2020, Feuillet_2021, Malhan_2022, Rix_2022, Malhan_2024, Tang_2024, Tian_2024, Zhang_2024}.

\citet{Yang_2019a} obtained more than 13,000 K-giant stars in the Galactic halo from LAMOST DR5 and Gaia DR2. Using the six-dimensional (6D) position-velocity information ($l$, $b$, $d$, $V_{\mathrm{los}}$, $V_{l}$, $V_{b}$) and the FoF clustering algorithm, substructures such as the Sgr Stream, Monoceros Ring, Virgo Overdensity, Hercules-Aquila Cloud, and Orphan Stream were identified. \citet{Naidu_2020} constructed a sample of 5,684 giant stars in the region $|b| > 40^\circ$ and $d_{\mathrm{helio}} > 3$ kpc from the H3 (“Hectochelle in the Halo at High Resolution”) Survey. Based on stellar properties in the IoM and action spaces, the analysis revealed various components of the Milky Way, including the Sgr Stream, Aleph, High-$\alpha$ Disk, In-situ Halo, GES, Helmi Stream, Thamnos, High-Energy Retrograde Halo (Sequoia, Arjuna, I’itoi), Wukong, and the Metal-Weak Thick Disk (MWTD). \citet{Wang_2022} compiled 3,003 ab-type RR Lyrae stars from SDSS DR12, LAMOST DR6, and Gaia EDR3, located in the Galactic halo. Orbital parameters in the IoM space were employed together with the FoF clustering algorithm to identify substructures including the Sgr Stream, GES, Sequoia, Helmi streams, and three additional groups possibly associated with globular clusters NGC 5272, 6656, and 5024. \citet{Tang_2024} selected 6,454 FGKM-type halo stars ($T_{\mathrm{eff}} = 3,500$–$8,000$ K, $\left| V - V_{\mathrm{LSR}} \right| > 210~\mathrm{km\,s^{-1}}$) from the medium-resolution data of LAMOST DR9. Using energy and angular momentum information in the IoM space, the HDBSCAN clustering algorithm enabled the identification of GES, Helmi streams, the high-$\alpha$ disk, the in-situ halo, and three additional clusters not clearly linked to known substructures.

While these studies primarily utilized K giants, BHBs, RR Lyrae stars, and other tracers to probe the Galactic halo, M giant stars—with their unique combination of high luminosity and lower susceptibility to interstellar extinction—offer complementary advantages for mapping both distant and low-latitude substructures. M giants are red giant branch (RGB) stars with low temperature and high luminosity \citep[][]{Li_2023}. On the one hand, the high luminosity allows them to be used to trace the distant volumes, which are in the Galactic halo. On the other hand, the lower temperature indicates that most of the flux is distributed at the long wavelength, which makes the M giants suffer lower susceptibility to extinction. This provides an opportunity to study the outer volumes of the disk with low latitude \citep[][]{Qiu_2023}. 

\citet{Li_2019} obtained an M giant star sample with complete phase-space information by combining LAMOST DR4 spectroscopic data, ALLWISE photometry, and accurate proper motions from Gaia DR2. Based on the typical kinematic properties of the Sgr stream, they selected 164 pure Sgr stream M giant stars. In the same year, \citet{Yang_2019b} obtained a large sample of M giants, K giants, and BHB stars by combining data from LAMOST DR5, SEGUE-2, 2MASS, and WISE. Using the FoF clustering algorithm in IoM space, they identified more than 3,000 candidate stars of the Sgr stream and provided a detailed characterization of their kinematic and chemical properties. Based on the spectroscopic data from LAMOST DR5 and the precise proper motions from Gaia DR2, \citet{Li_2021} applied the same method and identified 280 M giant candidates belonging to an anti-Galactic center substructure in the low-latitude region. These studies demonstrated the potential of M giants in tracing distant and low-latitude Galactic features.

The LAMOST DR9 catalog, with its expanded sample of 58,076 M-giant stars \citep{Li_2023, Qiu_2023}, provides a valuable resource for exploring substructures in the Galactic halo and disk. By combining Gaia DR3 data and applying FoF clustering methods in IoM space, we aim to identify substructures and study their properties.

This paper is organized as follows. Section \ref{sect:Data} describes the M giant sample used in this study. Section \ref{sect:Method} introduces the methods employed to identify substructures. In Section \ref{sect:Res}, we present the substructures identified through clustering and their characteristics. Finally, the discussion and conclusion are presented in Section \ref{sect:Discussion and Conclusion}.

\section{Data}\label{sect:Data}
In this work, we use the M giant sample from \citet{Li_2023} to investigate Galactic substructures. Based on the low-resolution spectra of LAMOST DR9, \citet{Li_2023} selected a reliable sample of 58,076 M giant stars and determined their radial velocities using the cross-correlation-based \emph{laspec} algorithm \citep{Zhang_2021}. The velocity offset and dispersion relative to Gaia DR3 and APOGEE DR17 are approximately 1 km s$^{-1}$ and 4.6 km s$^{-1}$, respectively. Based on this dataset, \citet{Qiu_2023} constrained the K-band absolute magnitude to estimate distances for the sample, achieving a relative statistical uncertainty of approximately $25\%$.

Combining the above data with precise proper motions provided by Gaia DR3 \citep{Gaia_2023}, we construct a catalog of 43,923 M giant stars, including their right ascension (R.A.), declination (Dec.), distances, radial velocities, and proper motions. Additionally, to investigate the origins of the substructures, we incorporate $[\mathrm{M}/\mathrm{H}]$ and $[\alpha/\mathrm{M}]$ parameters by cross-matching with APOGEE DR17 \citep{Abdurro'uf_2022}.

This work adopts both a Galactocentric Cartesian coordinate system \((X, Y, Z)\) and a Galactocentric spherical coordinate system \((r, \theta, \phi)\), where the positive direction of \(V_\phi\) is opposite to the rotation of Galactic disk. Using the Sun’s position of \((-8, 0, 0)\) kpc \citep{Reid_1993}, the solar motion of \((11.1, 12.24, 7.25)\) km s\(^{-1}\) \citep{schoenrich_2010} and the local standard of rest velocity of \(220\) km s\(^{-1}\) \citep{Kerr_1986}, we calculate the three-dimensional positions $(X_{gc}, Y_{gc}, Z_{gc})$ and velocities \((U, V, W)\) of all M giant stars in the Cartesian coordinate system, as well as the velocities \((V_r, V_\theta, V_\phi)\) in the spherical coordinate system.

\begin{figure*}[!ht]
\centering
\includegraphics[width=0.5\textwidth]{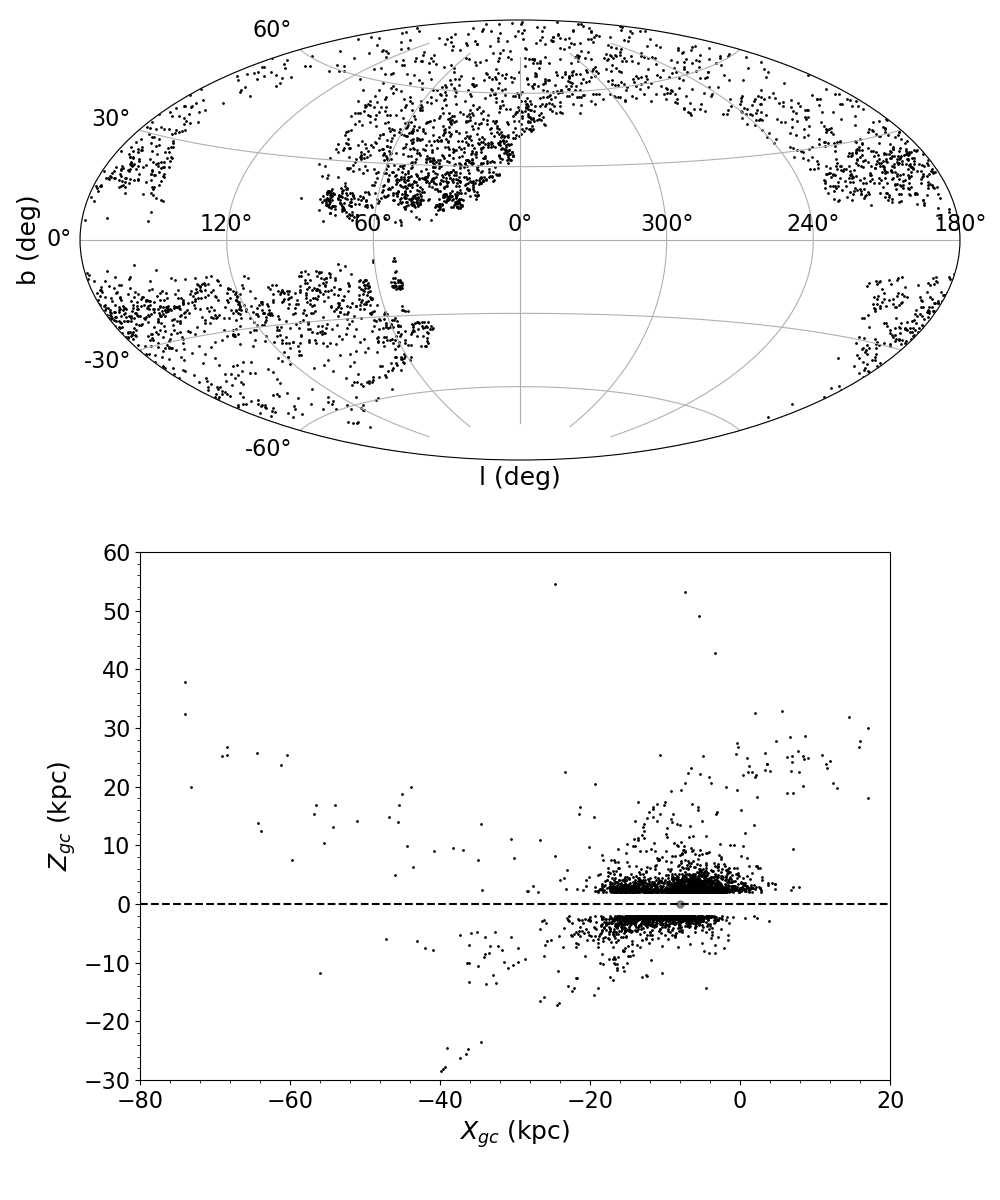}
\caption{\small The distribution of the selected M giant sample in phase space: Galactic coordinates (l, b) (upper panel) and $X_{gc}-Z_{gc}$ plane (lower panel).
}\label{fig:Sample1}
\end{figure*}

\begin{figure*}[!ht]
\centering
\includegraphics[width=0.8\textwidth]{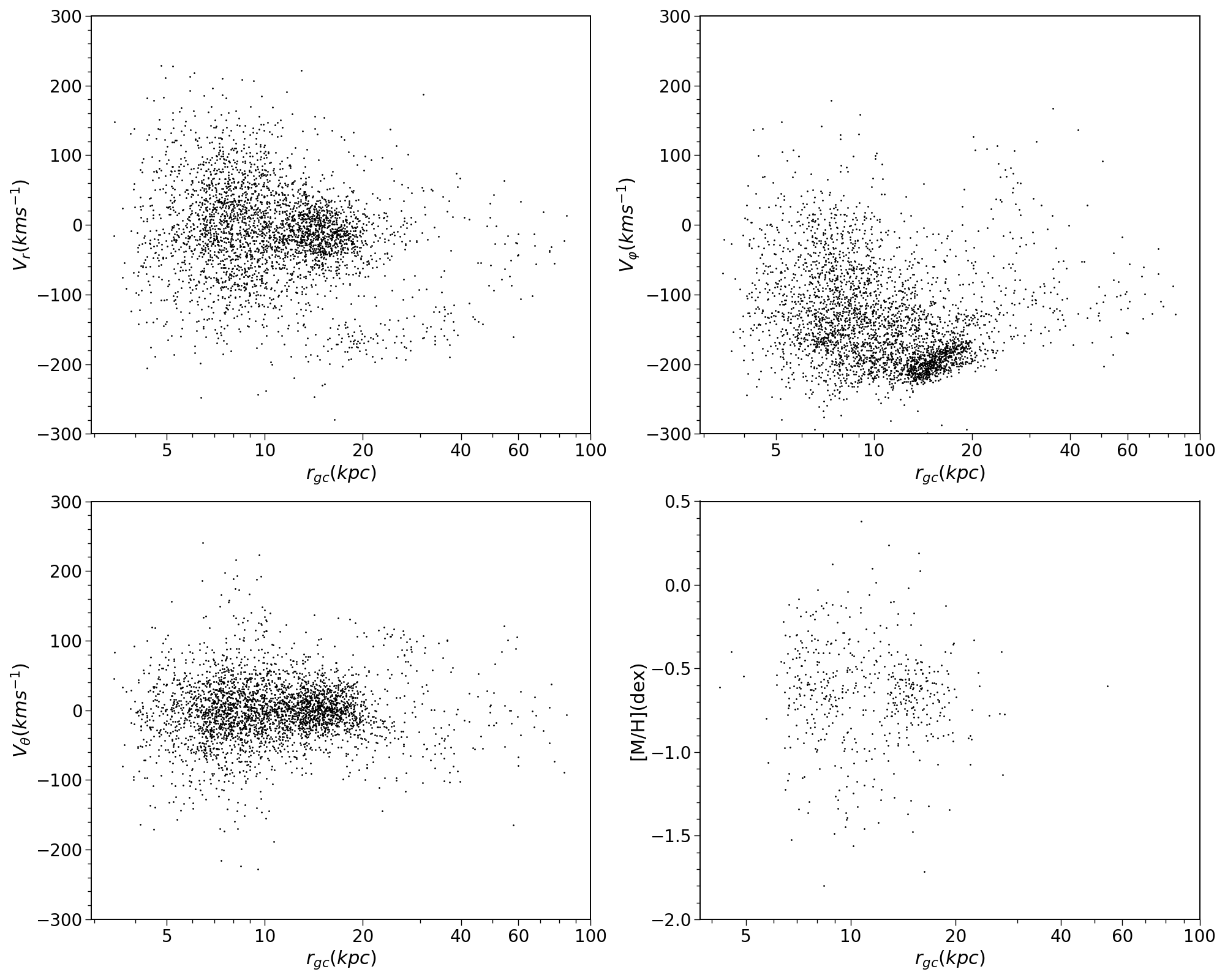}
\caption{\small The distribution of the selected M giant sample's velocities (V$_{\mathrm{r}}$, V$_{\mathrm{\phi}}$, V$_{\mathrm{\theta}}$) and $[\mathrm{M}/\mathrm{H}]$ relative to the Galactocentric distance (r$_{\mathrm{gc}}$).}
\label{fig:Sample2}
\end{figure*}

Considering the substantial number of disk stars in our sample, to minimize the impact of the thin disk and thick disk, we only include samples with \(|Z_{\text{gc}}| > 2\) kpc in the clustering analysis. In future work, we will further analyze the M giant sample with \(|Z_{\text{gc}}| < 2\) kpc. Additionally, to ensure data reliability, we retain only stars with a total velocity \( V_{\text{tot}} = \sqrt{U^2 + V^2 + W^2} \) less than \( 400 \, \text{km} \, \text{s}^{-1} \), and velocity component uncertainties \( \sigma_U \), \( \sigma_V \), and \( \sigma_W \) all smaller than \( 100 \, \text{km} \, \text{s}^{-1} \). As a result, the final sample used for identifying substructures consists of 3,343 M giant stars. The parameters of these stars are listed in Table~\ref{Table 1}. Figure~\ref{fig:Sample1} shows the spatial distribution of the sample in the Galactic coordinate system and the $X_{gc}$-$Z_{gc}$ plane. Furthermore, the velocity distributions (V$_{\rm r}$, V$_{\rm \phi}$, V$_{\rm \theta}$) and the metallicity distributions are shown in Figure~\ref{fig:Sample2}. Since M giants lack reliable chemical abundance parameters, we used K giants from \citet{Yang_2019b} as the disk star sample. These stars were cross-matched with APOGEE DR17 to derive $[\mathrm{M}/\mathrm{H}]$ and $[\alpha/\mathrm{M}]$ for comparison with the substructures.

\begin{table}[!ht]
    \centering
    \caption{\small Parameters of all M giant stars used for identifying Galactic substructures.}
    \resizebox{1\textwidth}{!}{ 
    \begin{tabular}{ccccccccccccccccc}
    \hline
        LAMOST & Gaia & APOGEE & R.A. & Dec. & d & $\text{d}_{\text{err}}$ & rv & $\text{rv}_{\text{err}}$ & pmra & $\text{pmra}_{\text{err}}$ & pmdec & $\text{pmdec}_{\text{err}}$ & [M/H]\textsuperscript{a} & [M/H]\textsubscript{err} & [\(\alpha\)/M]\textsuperscript{a} & [\(\alpha\)/M]\textsubscript{err}  \\ \hline
        783703241 & 6.02087E+17 & 2M08413271+1143476 & 130.386 & 11.7299 & 4.83058802 & 0.74150176 & 68.94289 & 0.10528198 & -1.041 & 0.021 & 0.354 & 0.016 & -0.55591 & 0.008231559 & 0.107612 & 0.00657604  \\ 
        158703190 & 2.66032E+18 & - & 351.848 & 4.87693 & 2.53512863 & 0.389145655 & -76.3728 & 0.10636841 & 8.169 & 0.018 & -8.97 & 0.013 & - & - & - & -  \\ 
        397903189 & 1.56292E+18 & 2M13232252+5232106 & 200.844 & 52.5362 & 2.398832919 & 0.36822408 & -85.57615 & 0.15323456 & -0.504 & 0.016 & -4.96 & 0.021 & -0.74163 & 0.008905697 & 0.261675 & 0.007471177  \\ 
        882612040 & 3.739E+18 & - & 203.599 & 12.3667 & 2.558585887 & 0.392746376 & -6.612488 & 0.10685303 & 1.117 & 0.036 & -16.564 & 0.018 & - & - & - & -  \\ 
        397804118 & 8.4228E+17 & 2M11070958+5208002 & 166.79 & 52.1334 & 2.582260191 & 0.396380413 & -76.66522 & 0.14766236 & 0.33 & 0.012 & -2.414 & 0.011 & -0.60106 & 0.008349871 & 0.123219 & 0.006826421  \\ 
        841612098 & 2.56403E+18 & - & 17.3769 & 4.64255 & 2.779713268 & 0.426689726 & -34.434998 & 0.11698608 & 3.149 & 0.027 & -6.372 & 0.018 & - & - & - & -  \\ 
        903102026 & 3.86886E+18 & - & 160.988 & 8.68141 & 2.60615355 & 0.400048076 & 19.425537 & 0.113559574 & -5.642 & 0.018 & 1.054 & 0.016 & - & - & - & -  \\ 
        906510249 & 1.44416E+18 & 2M13503093+2433256 & 207.629 & 24.5571 & 2.60615355 & 0.400048076 & -32.06021 & 0.10935791 & 0.231 & 0.034 & -0.869 & 0.025 & -0.32153 & 0.0077390824 & 0.092155 & 0.005329758  \\ 
        782303210 & 6.02275E+17 & 2M08432673+1239418 & 130.861 & 12.6616 & 5.152286446 & 0.790882901 & 15.706909 & 0.17940918 & 0.933 & 0.016 & -5.709 & 0.012 & -1.2221 & 0.010783301 & 0.15455501 & 0.010562286  \\ 
        442215130 & 1.51355E+18 & - & 190.663 & 31.3892 & 2.937649652 & 0.450933173 & -11.605164 & 0.1572583 & -11.118 & 0.024 & -8.402 & 0.025 & - & - & - & -  \\ \hline
    \end{tabular}
    } 
    \label{Table 1}
    \begin{flushleft}
    \textbf{Notes.} \\
    \textsuperscript{a} $[\mathrm{M}/\mathrm{H}]$ and $[\alpha/\mathrm{M}]$ from APOGEE DR17. \\
    (This table is available in its entirety in machine-readable form.)
    \end{flushleft}
\end{table}

\section{Method}\label{sect:Method}
Over its evolutionary history, particularly in the first three billion years, the Milky Way has undergone multiple mergers with dwarf galaxies. Dynamical theory predicts that, despite the loss of spatial coherence from dynamical mixing, orbital characteristics persist in IoM space. This preservation offers a promising approach to trace the properties of their progenitor systems (\citealt{Roederer_2018, Massari_2019, Koppelman_2019a, Malhan_2022, Malhan_2024}; Xue et al. 2024, in preparation). Recent studies (e.g., \citealt{Yang_2019b, Koppelman_2019b, Naidu_2020, Li_2021, Wang_2022, Tang_2024, Zhang_2024, Sun_2024}) have demonstrated the feasibility of identifying substructures in IoM space.

In a spherical potential without considering dynamic friction, Xue et al. (2024, in preparation) defined five IoM parameters: ($ec$, $a$, $l_{\mathrm{orbit}}$, $b_{\mathrm{orbit}}$, $l_{\mathrm{apo}}$) \citep{Yang_2019b, Li_2021, Wang_2022}. The definitions of the parameters are as follows:

\begin{itemize}
    \item[\textbullet] \textbf{Orbital eccentricity $ec$:} $ec = \frac{r_{\mathrm{apo}} - r_{\mathrm{per}}}{r_{\mathrm{apo}} + r_{\mathrm{per}}}$,
    \item[\textbullet] \textbf{Semi-major axis $a$:} $a = \frac{r_{\mathrm{apo}} + r_{\mathrm{per}}}{2}$,
    \item[\textbullet] \textbf{Orbital pole direction $(l_{\mathrm{orbit}}, b_{\mathrm{orbit}})$:}  
    $l_{\mathrm{orbit}} = \arctan\left(\frac{L_y}{L_x}\right)$ and 
    $b_{\mathrm{orbit}} = \arcsin\left(\frac{L_z}{L}\right)$,
    \item[\textbullet] \textbf{Apocenter direction $l_{\mathrm{apo}}$:} the angle between the apocenter and the projection of the $x$-axis on the orbital plane.
\end{itemize}

To quantify the orbital separations between stars, we define the orbit-likelihood distance $\delta_{ij}$ following \citet{Wang_2022}:

\begin{align}
\delta_{ij}^2 = & \, \omega_{\theta} \, \theta_{ij}^2 + \omega_{\Delta a} \, (a_i - a_j)^2 + \omega_{\Delta e} \, (e_i - e_j)^2 + \omega_{\Delta l_{\text{apo}}} \, (l_{\text{apo}, i} - l_{\text{apo}, j})^2
\end{align}

Here, $\omega_{\theta}$, $\omega_{\Delta a}$, $\omega_{\Delta e}$, and $\omega_{\Delta l_{\text{apo}}}$ are normalization weights (see \citealt{Wang_2022} for details). The FoF algorithm is employed for substructure identification. In the FoF algorithm, two stars are linked into a group if their orbit-likelihood distance $\delta$ is smaller than a specified threshold (linking length). Stars are subsequently added to the group if $\delta$ between them and any group member is smaller than the linking length. This process is iterated until no more stars can be added to the group. At this point, the group construction is complete \citep{Wang_2022, Sun_2024}.

To ensure that the members of each substructure are identified as completely as possible and obtain reliable clustering results, we set the linking length range from 0.138 to 0.487 and only retain groups containing more than 10 members. Details of the linking length selection are provided in Appendix \ref{APP:A}. The maximum physical sizes of each component corresponding to the linking length of 0.138 and 0.487 are summarized in Table~\ref{Table 2}.

\begin{table}[!ht]
    \centering
    \begin{tabular}{ccccc}
    \hline
    Linking length & $\theta(\mathrm{l_{orbit}}, \mathrm{b_{orbit}})$ (deg) & a (kpc) & e & $\ell_{\mathrm{apo}}$ (deg) \\ \hline
    0.138 & 7.36 & 1.46 & 0.04 & 13.90 \\ 
    0.487 & 25.99 & 5.17 & 0.16 & 49.05 \\ \hline
    \end{tabular}
    \caption{\small The maximum physical sizes of each component corresponding to the linking lengths of 0.138 and 0.487.
    }\label{Table 2}
\end{table}

\section{Results}\label{sect:Res}

\begin{figure*}[!ht]
\centering
\includegraphics[width=0.7\textwidth]{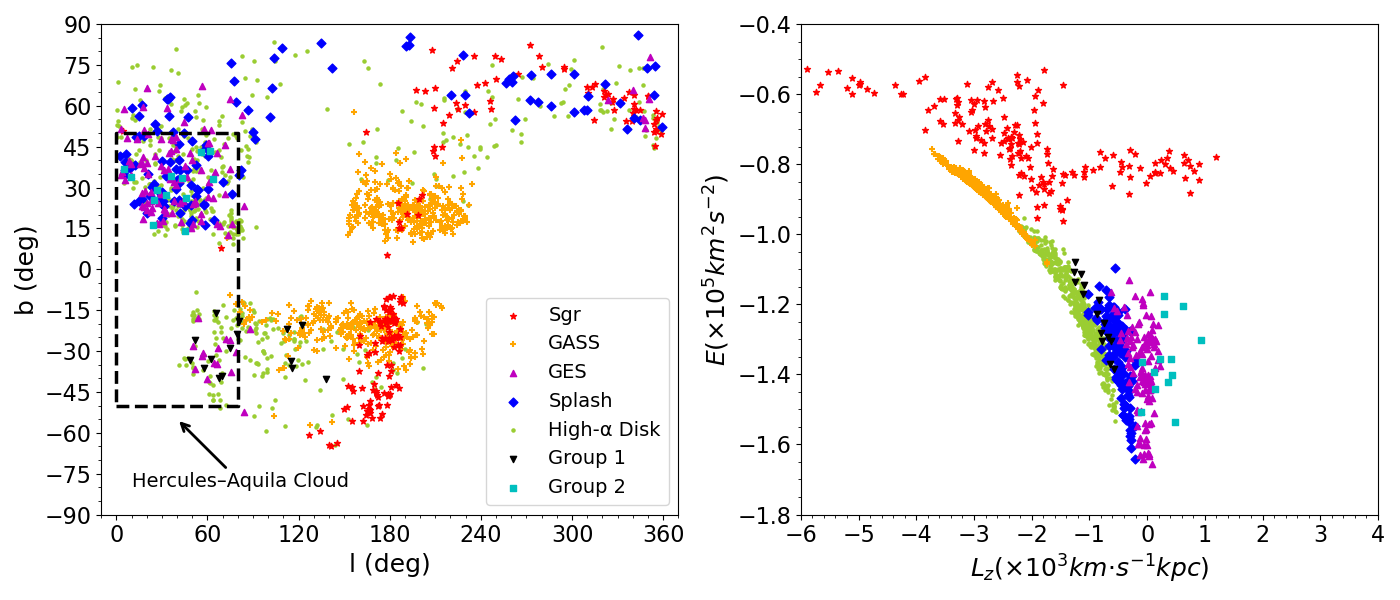}
\caption{\small The distributions of groups associated with known substructures in the Galactic coordinate system and the $L_{\mathrm{z}}$-E plane. The figure includes the known substructures Sgr (red pentagram), GASS (orange plus), GES (purple triangle), Splash (blue diamond), and High-$\alpha$ Disk (yellow-green dots), as well as two possible new substructures: Group 1 (black downward triangle) and Group 2 (cyan square). The black dashed areas in the left panel indicate the location of the Hercules–Aquila Cloud as defined by \citet{Belokurov_2007}.
}\label{fig:all substructures}
\end{figure*}

As described in Section \ref{sect:Method}, we apply the FoF algorithm to search for substructures in the IoM space. We ultimately identify 47 groups containing a total of 1,597 M giant stars. By comparing their spatial distribution, kinematics, dynamics, and chemical properties with those of known substructures, we find that the groups mainly belong to the Sgr stream \citep{Belokurov_2014, Hernitschek_2017, Yang_2019b}, Galactic Anticenter Substructure (GASS) \citep{Li_2021}, GES \citep{Belokurov_2018, Naidu_2020}, Splash \citep{Belokurov_2020}, and the high-$\alpha$ disk \citep{Naidu_2020, Tang_2024}. Figure~\ref{fig:all substructures} shows the distribution of these substructures in the Galactic coordinate system ($l$, $b$) and the energy-angular momentum ($L_{\mathrm{z}}$-$E$) space. Additionally, we find two new groups which are not associated with any known substructures. In the following, we will provide a detailed explanation of these substructures. Table~\ref{Table 3} shows the orbital parameters of all identified substructures.

\subsection{The Sagittarius stream}\label{sect:sgr}

The Sgr stream, the most prominent substructure in the Galactic halo, exhibits a growing population of spectroscopically confirmed member stars. This expanded census facilitates a comprehensive investigation of the dynamical interaction between the Sagittarius dwarf spheroidal galaxy (Sgr dSph) and the Milky Way \citep{Ibata_1994, Yanny_2000, Ibata_2001, Belokurov_2006, Carlin_2018, Li_2019, Yang_2019b}. By comparing our groups with previous observational Sgr members \citep{Belokurov_2014, Hernitschek_2017, Yang_2019b}, we identify three groups (182 stars) associated with the Sgr stream: one group (67 stars) in the Sgr leading arm, one group (105 stars) in the Sgr trailing arm, and one group (10 stars) in the Sgr debris. We cross-matched the Sgr sample with APOGEE DR17 and obtained 19 stars with reliable $[\mathrm{M}/\mathrm{H}]$ and $[\alpha/\mathrm{M}]$ measurements.

\begin{figure*}[!ht]
\centering
\includegraphics[width=0.7\textwidth]{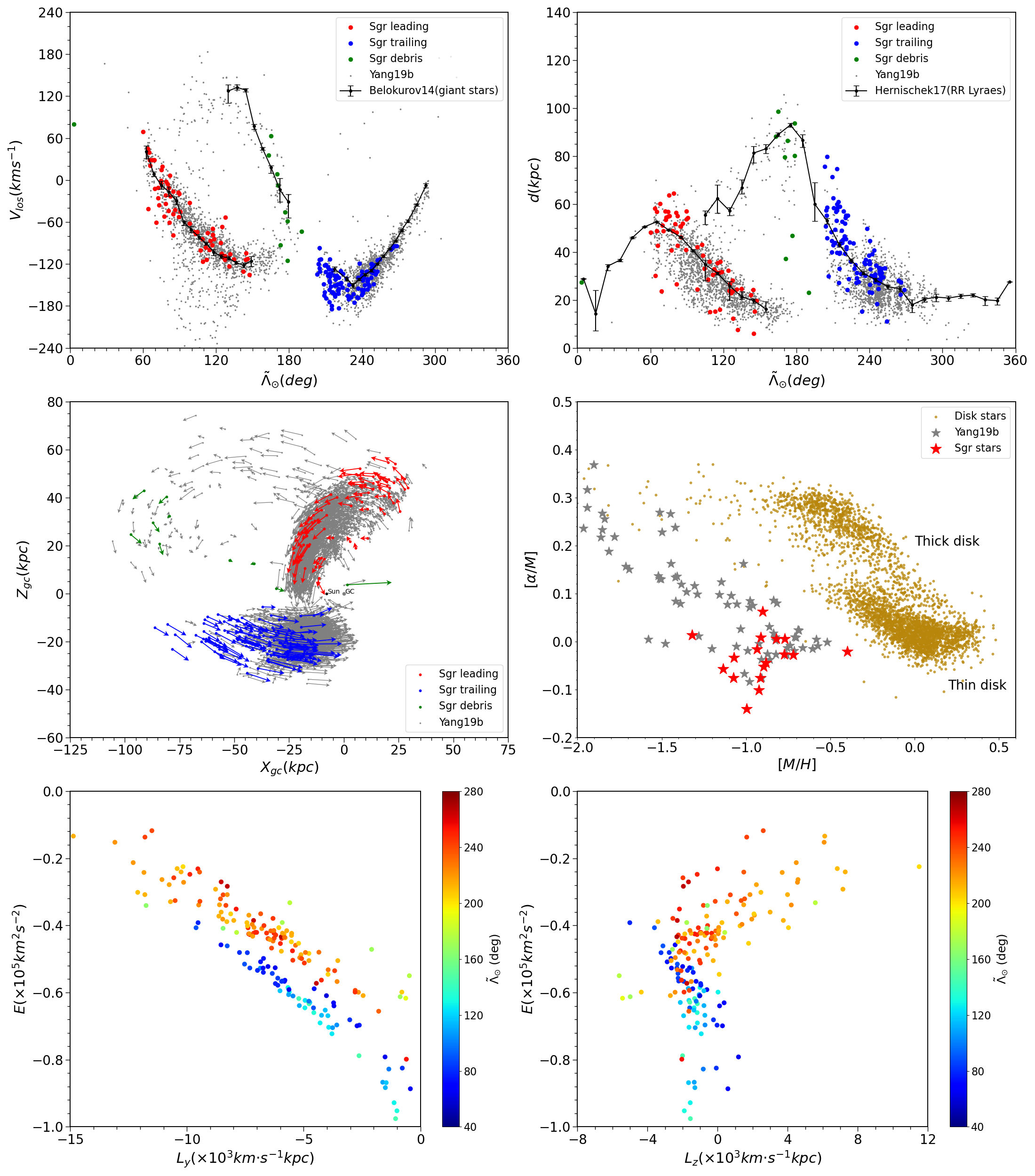}
\caption{\small The distributions of the Sgr members in the $\tilde{\Lambda}_{\odot}$-$V_{\mathrm{los}}$, $\tilde{\Lambda}_{\odot}$-$d$, $X_{\mathrm{gc}}$-$Z_{\mathrm{gc}}$, $[\mathrm{M/H}]$-$[\alpha/\mathrm{M}]$, $L_z$-$E$, and $L_y$-$E$ planes. $\tilde{\Lambda}_{\odot}$ is the longitude in the Sgr coordinate system and the definition is the same as that in \citet{Belokurov_2014}. The red, blue, and green points in the figure represent members of the Sgr leading arm, trailing arm, and debris, respectively, while the gray points represent the observational sample from \citet{Yang_2019b}. The black points with error bars in the upper-left panel represent the giant stars from \citet{Belokurov_2014}. The black points with error bars in the upper-right panel represent the RR Lyrae stars from \citet{Hernitschek_2017}. Arrows in the middle-left panel indicate the direction and amplitude of the velocity in the $X_{gc}$-$Z_{gc}$ plane. The middle-right panel shows a comparison of our Sgr stars (red stars) with Galactic disk stars (yellow points) and Sgr stars from \citet{Yang_2019b} in the $[\alpha/\mathrm{M}]$ vs. $[\mathrm{M/H}]$ space. All $[\mathrm{M}/\mathrm{H}]$ and $[\alpha/\mathrm{M}]$ measurements are taken from APOGEE DR17. The lower two panels show energy-angular momentum distribution and the color indicates the $\tilde{\Lambda}_{\odot}$ value.
}\label{fig:Sgr}
\end{figure*}

\citet{Qiu_2023} estimated the distances of M giant stars based on their intrinsic color index, metallicity, and extinction. However, \citet{Li_2016a} suggested that the distances of M giants are not solely determined by these parameters but are also strongly affected by the star formation history of their host system. Consequently, compared to Sgr member stars identified using other stellar tracers, the distances derived by \citet{Qiu_2023} for the Sgr stream are noticeably underestimated \citep{Belokurov_2014, Hernitschek_2017, Yang_2019b}. To address this, we correct the distances of the identified Sgr member stars using the color–absolute magnitude relation derived from \citet{Li_2016a}. Figure~\ref{fig:Sgr} presents a comparison of our Sgr stream sample with other observational data \citep{Belokurov_2014, Hernitschek_2017, Yang_2019b} across the ($\tilde{\Lambda}_{\odot}$, $V_{\mathrm{los}}$), ($\tilde{\Lambda}_{\odot}$, $d$), and ($X_{\mathrm{gc}}$, $Z_{\mathrm{gc}}$) spaces. The velocities and distances of our identified Sgr members show excellent agreement with those from other observational studies.

From the spatial distribution in the $X_{gc}$-$Z_{gc}$ plane and energy-angular momentum distribution shown in Figure~\ref{fig:Sgr}, the traditionally identified leading stream appears to consist of two distinct orbital components: one part vertically passes through the Galactic disk, while the other part runs almost parallel to the disk. In the energy-angular momentum distribution, we also see that the leading members are clearly divided into dark blue and light blue groups based on their $\tilde{\Lambda}_{\odot}$ values. Further work will focus on a more in-depth orbital analysis by incorporating different types of Sgr member stars.

Additionally, in the middle-right panel of Figure~\ref{fig:Sgr}, our Sgr stars exhibit a distribution sequence similar to that of the Sgr stars identified in K giants, M giants and BHBs from \citet{Yang_2019b}, characterized by lower $[\alpha/\mathrm{M}]$ abundances compared to Galactic disk stars at high metallicity. This chemical evolution pattern is consistent with Milky Way dwarf galaxies \citep{Venn_2012, Lemasle_2014, Yang_2019b}.

\subsection{The Galactic Anticenter Substructure}\label{sect:Gass}

The GASS is a substructure located at low Galactic latitudes and in the anti-center direction of the Milky Way. Due to their overlapping kinematics and spatial properties, \citet{Li_2021} grouped several previously identified anti-center substructures, including Monoceros \citep{Newberg_2002, Li_2012}, A13 \citep{Sharma_2010, Li_2017}, and the Triangulum-Andromeda cloud \citep{Rocha-Pinto_2004, Deason_2014}. \citet{Li_2021} obtained a GASS sample containing 589 stars, among which 280 stars are M giants, and suggested that GASS is part of the local metal-poor outer disk of the Milky Way.

\begin{figure*}
\centering
\includegraphics[width=0.9\textwidth]{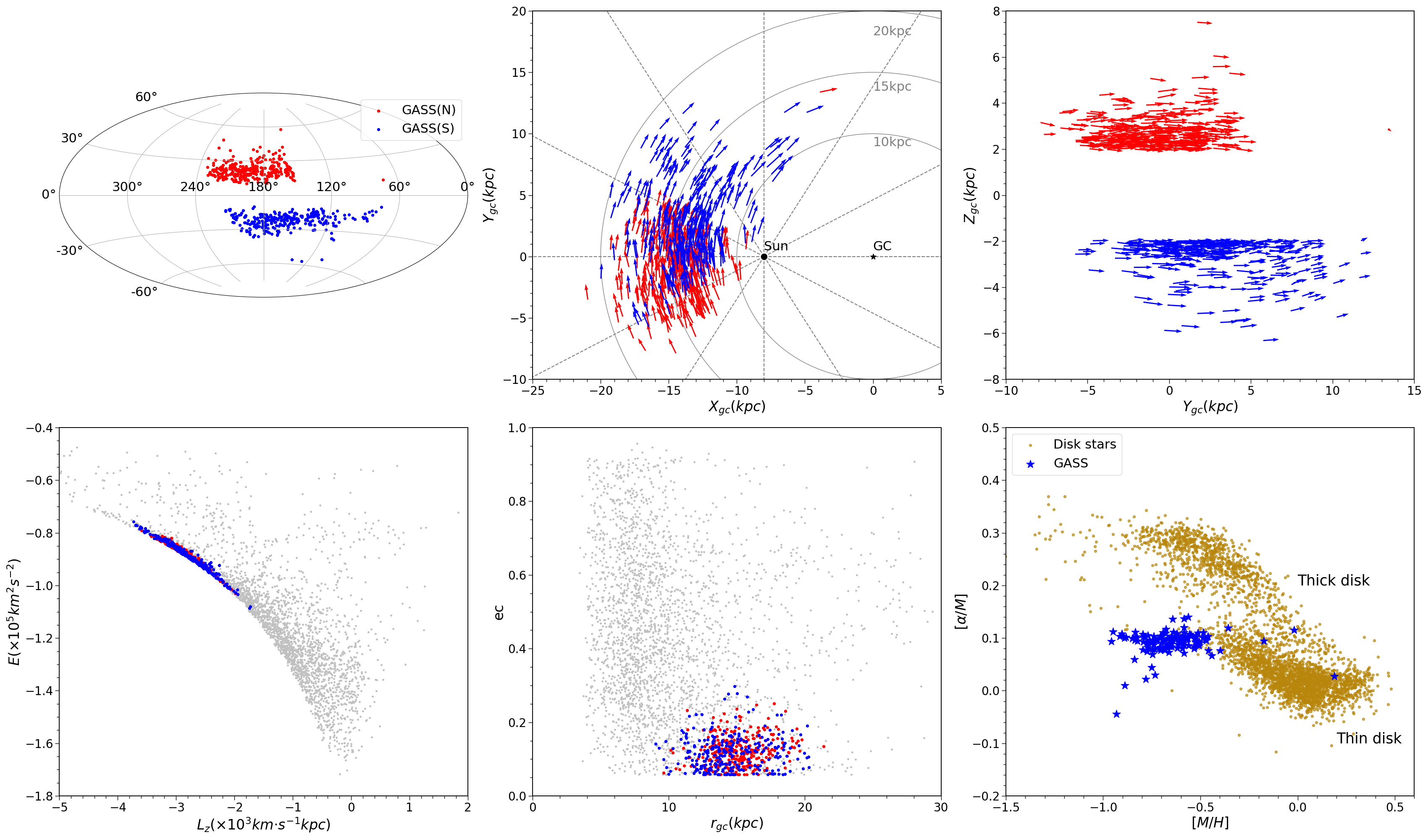}
\caption{\small Distribution of our GASS members in $(l, b)$, $(X_{\mathrm{gc}}, Y_{\mathrm{gc}})$, $(Y_{\mathrm{gc}}, Z_{\mathrm{gc}})$, $(L_{\mathrm{z}}, E)$, $(r_{\mathrm{gc}}, ec)$, and $([\mathrm{M}/\mathrm{H}], [\alpha/\mathrm{M}])$ spaces. The blue and red points represent the members in the northern and southern hemispheres respectively. The silver points represent all M giant stars. In the second and third panels of the upper row, arrows indicate the direction and amplitude of the velocity in the $X_{\mathrm{gc}}$-$Y_{\mathrm{gc}}$ and $Y_{\mathrm{gc}}$-$Z_{\mathrm{gc}}$ planes. In the third panel of the lower row, the blue stars represent 122 GASS M giants, while the yellow points represent the disk stars from \citet{Yang_2019b}. All [$[\mathrm{M}/\mathrm{H}]$] and $[\alpha/\mathrm{M}]$ measurements are taken from APOGEE DR17.}
\label{fig:GASS1}
\end{figure*}

By comparing the properties of the GASS with those reported in \citet{Li_2021}, we identify 11 groups comprising a total of 594 M giant stars associated with the GASS. Among these, 313 stars are located in the northern part of the Galactic disk, and 281 stars reside in the southern part. We cross-matched the GASS sample with APOGEE DR17 and obtained 122 stars with reliable $[\mathrm{M}/\mathrm{H}]$ and $[\alpha/\mathrm{M}]$ measurements. Figure~\ref{fig:GASS1} shows the distribution of the GASS members in $(l, b)$, $(X_{\mathrm{gc}}, Y_{\mathrm{gc}})$, $(Y_{\mathrm{gc}}, Z_{\mathrm{gc}})$, $(L_{\mathrm{z}}, E)$, and $(r_{\mathrm{gc}}, ec)$ spaces, as well as the comparison between GASS and disk stars in the $[\mathrm{M}/\mathrm{H}]$–$[\alpha/\mathrm{M}]$ planes. In the upper left panel, it can be seen that these members cover a large region at low latitude with a range of \(100^\circ < l < 240^\circ\) and \(-40^\circ < b < 40^\circ\). In the $X_{gc}$-$Y_{gc}$ plane, the GASS exhibits a prograde circular orbit. Additionally, in the $Y_{gc}$-$Z_{gc}$ plane, the GASS has a slight arc motion trend, particularly in the southern hemisphere, consistent with the results of \citet{Li_2021}. \citet{Li_2021} suggested that it could be part of the expected ripple pattern in \citet{Xu_2015} and \citet{Li_2017}. In the $L_{\mathrm{z}}$-$E$ plane, it is located in an extended narrow region of thin disk stars, but has a higher \(E\) and higher \(L_{\mathrm{z}}\) value than most of the thin disk population. The $r_{\mathrm{gc}}$-$ec$ plane shows that the GASS mainly distributes in the range of \(r_{\mathrm{gc}} = 10 \, \mathrm{kpc}\) to \(20 \, \mathrm{kpc}\), with an orbital eccentricity \(ec < 0.2\), indicating its very circular orbit.

In the lower-right panel of Figure~\ref{fig:GASS1}, the comparison with disk stars shows that GASS has similar $\alpha$-element abundances to the thin disk, but is more metal-poor than typical thin disk stars. This feature is consistent with the continuation of metal-rich thin disk stars into the outer halo \citep{Haywood_2016, Hayes_2018, Li_2021}. Our GASS members share a similar distribution in the $[\alpha/\mathrm{M}]$ versus $[\mathrm{M}/\mathrm{H}]$ space with those identified by \citet{Li_2021}. According to \citet{Li_2021}, GASS may have formed in the outer disk region after the thick disk phase, where lower molecular cloud densities resulted in a reduced star formation efficiency compared to the inner disk.

\subsection{Gaia-Enceladus-Sausage}

The GES is a merger remnant in the Milky Way, formed through a major accretion event with a dwarf galaxy approximately 8–10 billion years ago \citep{Belokurov_2018, Koppelman_2018, Myeong_2018, Haywood_2018, Helmi_2018, Mackereth_2019}. \citet{Belokurov_2018} first identified the GES based on a large sample of main-sequence stars within $\sim10$ kpc, combining data from Gaia  DR1 and SDSS. They named it the “Gaia Sausage” due to its distinctive sausage-like distribution in the $V_r$-$V_\phi$ plane, where $V_r$ is the radial velocity and $V_\phi$ is the azimuthal velocity,  with $V_\phi \sim 0 \, \text{km}~\text{s}^{-1}$ indicating highly radial orbits. Since its discovery, the GES has been extensively studied \citep{Koppelman_2018, Myeong_2018, Haywood_2018, Helmi_2018, Mackereth_2019, Naidu_2020, Zhao_2021, Wang_2022, Tang_2024}.

\begin{figure*}[!ht]
\centering
\includegraphics[width=0.7\textwidth]{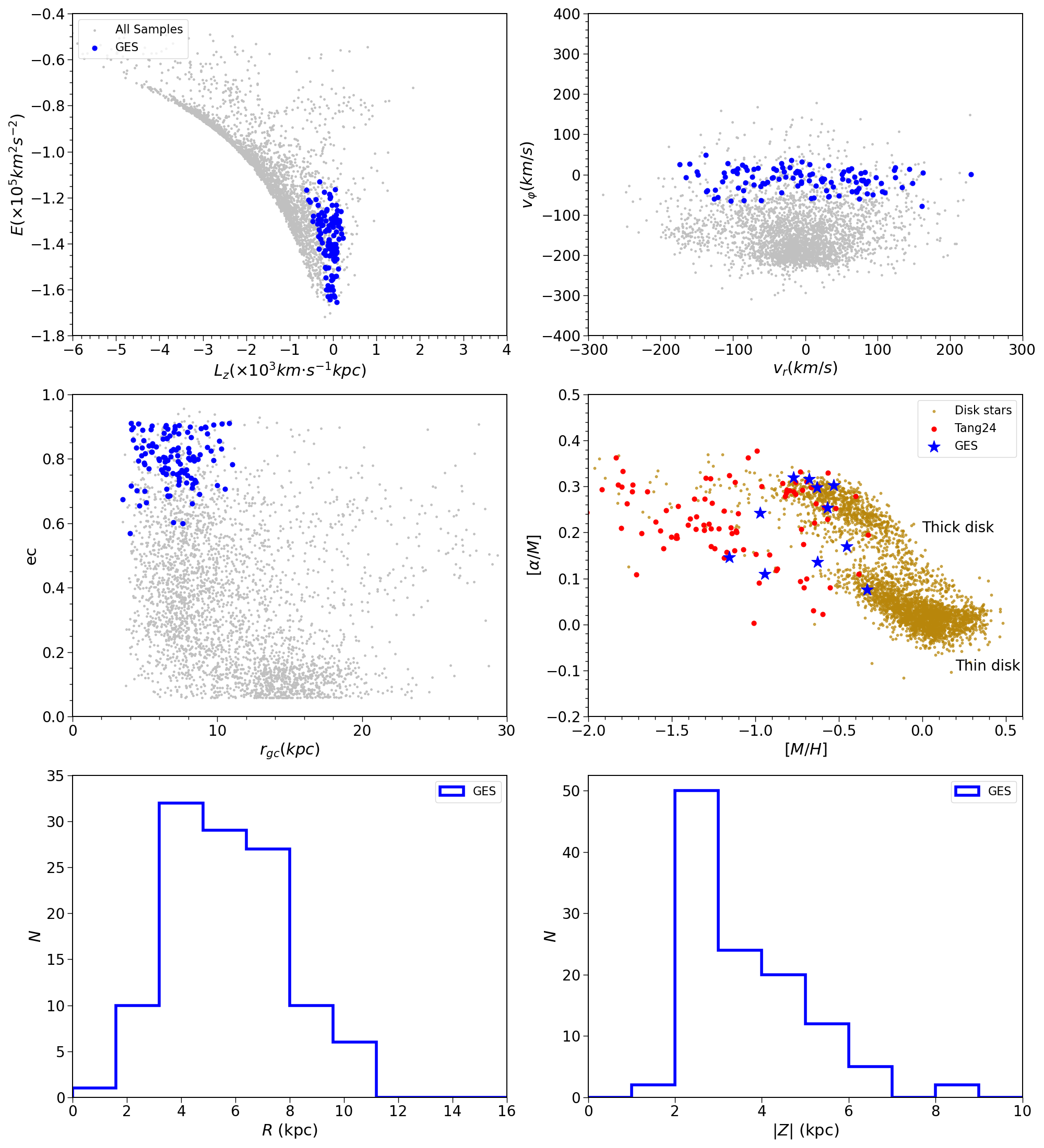}
\caption{\small Distribution of the GES in the $L_{\mathrm{z}}$-$E$, $V_{r}$-$V_{\phi}$, $r_{\mathrm{gc}}$-$ec$, and $[\mathrm{M}/\mathrm{H}]$-$[\alpha/\mathrm{M}]$ planes, as well as histograms of $R$ and $|Z|$. The blue points represent all the selected GES member stars, while the silver points represent all M giant stars. In the middle-right panel, the blue and red star symbols represent the GES M giants from our sample and the GES stars from \citet{Tang_2024}, respectively. The yellow points represent disk stars from \citet{Yang_2019b}. All $[\mathrm{M}/\mathrm{H}]$ and $[\alpha/\mathrm{M}]$ measurements are taken from APOGEE DR17. The lower two panels show the histograms of $R$ and $|Z|$ for our GES, respectively.
}\label{fig:GES1}
\end{figure*}

In our sample, we identify seven groups belonging to the GES, containing a total of 115 M giant stars. By cross-matching the GES sample with APOGEE DR17, we obtained 11 stars with reliable $[\mathrm{M}/\mathrm{H}]$ and $[\alpha/\mathrm{M}]$ measurements. As shown in Figure~\ref{fig:GES1}, our selected GES members are kinematically consistent with the typical GES. In the $(L_z, E)$ plane, the GES shows a vertical bar-like distribution near $L_z \sim 0 \, \text{kpc} \, \text{km} \, \text{s}^{-1}$. In the $(V_r, V_\phi)$ plane, it shows a sausage-like distribution near $V_\phi \sim 0 \, \text{km}~\text{s}^{-1}$ with a broad $V_r$ range  and a high eccentricity ($ec > 0.7$).

\citet{Zhao_2021} selected metal-rich Sausage-kinematic (MRSK) stars from the LAMOST DR5 dataset, defined by $[\mathrm{Fe}/\mathrm{H}] > -0.8$ dex and $-100 < V_\phi < 50$ km s$^{-1}$, then categorized them into low-$\alpha$ and high-$\alpha$ MRSK subgroups using their $\alpha$-element abundances. In a complementary study, \citet{Amarante_2020} conducted a hydrodynamical simulation of a clumpy Milky-Way-like analogue that successfully replicated bimodal disk chemistry with Sausage-like kinematics. Building on the consistency between MRSK stellar properties and these simulations, \citet{Zhao_2021} hypothesized that both low-$\alpha$ and high-$\alpha$ MRSK population originates from the GES progenitor galaxy. They posited that the ancient GES merger introduced gas-rich clouds into the Milky Way during its early evolution, subsequently forming clumps that generated the observed bimodal chemistry through later evolutionary processes.

As demonstrated in the middle-right panel of Figure~\ref{fig:GES1}, our metal-rich GES M giants display two distinct $\alpha$-abundance sequences (high-$\alpha$ vs. low-$\alpha$),  thereby corroborating \citet{Zhao_2021}'s scenario. Furthermore, comparative analysis with \citet{Tang_2024}'s GES stars reveals that our M giants trace the GES metallicity extension from metal-poor to metal-rich regimes, strengthening \citet{Zhao_2021}'s interpretation of MRSK stars as the metal-rich component of the GES. Finally, for our GES sample, the $R$ and $|Z|$ distributions ($2 < R < 12$ kpc; $2 < |Z| < 8$ kpc) align with the metal-rich GES spatial trends reported in Figure 5 of \citet{Zhao_2021}.

\begin{figure*}
\centering
\includegraphics[width=0.8\textwidth]{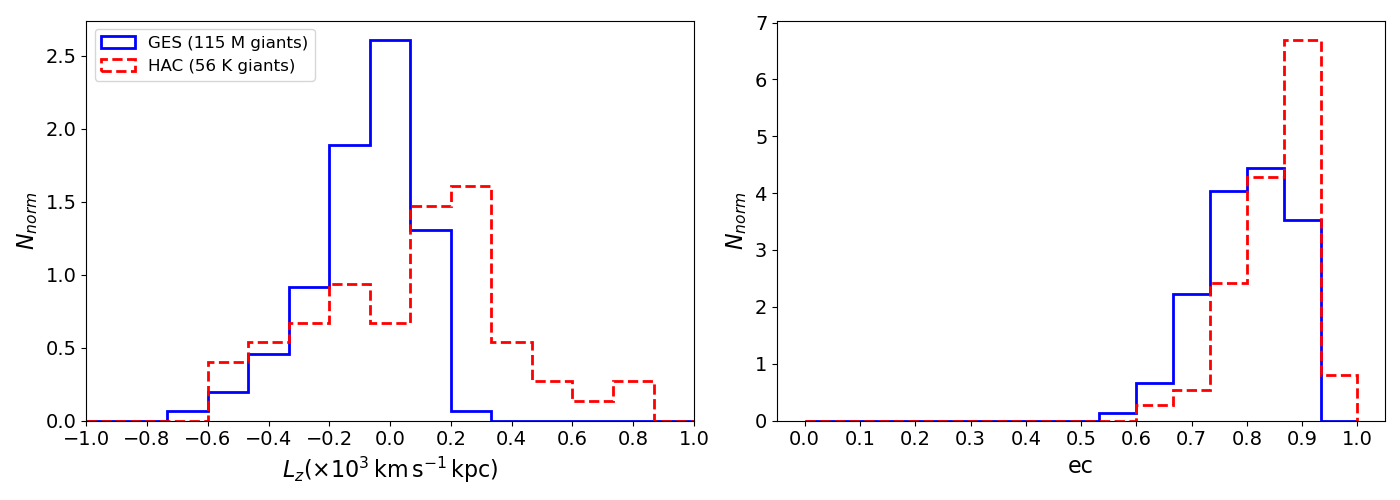}
\caption{\small The distribution of the angular momentum ($L_z$) and eccentricity ($ec$) of GES members. The blue solid line traces the distribution of the GES members in our sample, while the red dashed line traces the distribution of HAC K giants from \citet{Yang_2019a}.
}\label{fig:GES2}
\end{figure*}

Previous studies \citep[e.g.][]{Li_2016b, Balbinot_2021, Wang_2022} have suggested a connection between GES and the Hercules–Aquila Cloud (HAC). \citet{Belokurov_2007} first discovered the HAC and suggested that it covers a vast area of the sky, centered at a Galactic longitude of $l \sim 40^\circ$, with Galactic latitude $b$ from $ -50^\circ$ to $+50^\circ$. Figure\ref {fig:all substructures} shows that most of our GES member stars are located within the HAC region. Therefore, to explore their connection, we compare the GES with the HAC K-giant sample identified by \citet{Yang_2019a}. As shown in Figure~\ref{fig:GES2}, the GES, similar to the HAC, exhibits high orbital eccentricity and low $L_z$. The slight deviation between the GES members and HAC K giants in the $L_z$ distribution can be attributed to the insufficient number of M giants. Given similar kinematic properties of GES and HAC, we reach the same conclusion as \citet{Wang_2022} that GES and HAC may have similar origins.

\subsection{Splash}
The existence of the in situ halo in the Milky Way has long been a topic of significant interest. \citet{Bonaca_2017} constructed a large sample of high-eccentricity stars with $[\mathrm{Fe}/\mathrm{H}] > -1$ using Gaia DR1 data. Based on chemical abundance analysis and comparisons with numerical simulations, \citet{Bonaca_2017} argued that these locally observed metal-rich stars on halo-like orbits were most likely formed within the Galaxy. Subsequent studies based on Gaia DR2 data have also identified stellar populations that exhibit both high orbital eccentricities and thick disk chemical abundance patterns, and have referred to one such component as the “heated thick disc”, interpreted as the in situ halo \citep{Haywood_2018, Matteo_2019, Amarante_2020}.

In particular, \citet{Belokurov_2020} confirmed the existence of a large metal-rich population of stars ($[\mathrm{Fe}/\mathrm{H}] > -0.7$) on highly eccentric orbits ($ec > 0.5$) in the solar neighborhood. Their analysis leveraged Gaia DR2 astrometry and spectroscopic data for bright nearby stars. By analyzing the evolution of kinematics, elemental abundances and stellar ages in the $V_\phi$–$[\mathrm{Fe}/\mathrm{H}]$ plane, \citet{Belokurov_2020} further proposed that these stars likely originated in the Milky Way’s proto-disk prior to a massive ancient accretion event, such as the GES merger which drastically perturbed their orbits. Thus, this component was named “Splash”. This scenario provides crucial insights into the early dynamical evolution of the Milky Way. Follow-up studies have since characterized Splash stars via their distinct phase-space signatures in the energy–angular momentum space and chemical abundance patterns \citep{Naidu_2020, Bonaca_2020, Myeong_2022, Deepak_2024, Tang_2024}.

\begin{figure*}
\centering
\includegraphics[width=0.7\textwidth]{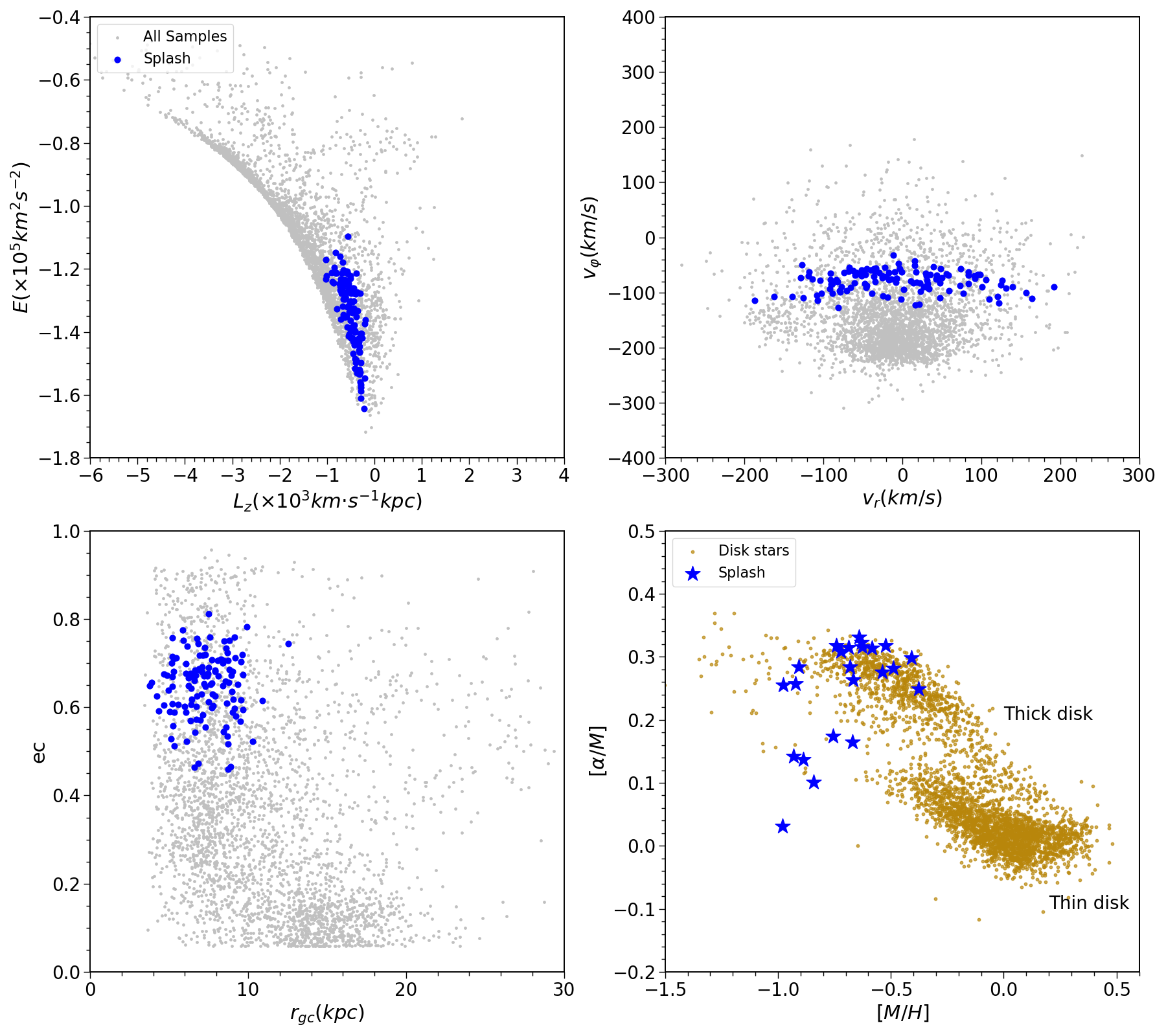}
\caption{\small Distribution of the Splash in the $L_{\mathrm{z}}$-$E$, $V_{r}$-$V_{\phi}$, $r_{\mathrm{gc}}$-$ec$, and $[\mathrm{M}/\mathrm{H}]$-$[\alpha/\mathrm{M}]$ planes. The blue points and silver points represent the Splash members and all M giants, respectively. In the bottom-right panel, the blue stars represent 23 Splash M giants, while the yellow points represent the disk stars from \citet{Yang_2019b}. All $[\mathrm{M}/\mathrm{H}]$ and $[\alpha/\mathrm{M}]$ measurements are taken from APOGEE DR17.
}\label{fig:Splash1}
\end{figure*}

In this work, we identify seven groups (121 stars) as members of the Splash component and characterize their kinematic and chemical properties, such as high orbital eccentricities ($ec > 0.5$) and thick disk chemistry ($[\mathrm{M}/\mathrm{H}] > -1$ dex) \citep{Belokurov_2020, Naidu_2020, Bonaca_2020, Deepak_2024}. By cross-matching the Splash sample with APOGEE DR17, we obtained 23 stars with reliable $[\mathrm{M}/\mathrm{H}]$ and $[\alpha/\mathrm{M}]$ measurements.

As shown in Figure~\ref{fig:Splash1}, Splash candidates occupy angular momenta $-1 \times 10^3 < L_z < 0 \, \text{kpc} \, \text{km}~\text{s}^{-1}$, yet their \(L_z\) values are systematically lower than thick disk stars at matched energies. In the \(V_r\)-\(V_\phi\) plane, Splash members predominantly lie above the kinematic boundary \(V_\phi \sim -100\,\mathrm{km\,s^{-1}}\) \citep{Belokurov_2020}, distinguishing them from the thick disk. Additionally, their \(V_r\) distributions show significantly greater dispersion compared to the thick disk population. As shown in the lower-left panel of Figure~\ref{fig:Splash1}, Splash stars exhibit distinctly high orbital eccentricities ($ec > 0.5$), differentiating them from thick disk stars. However, in the chemical plane of $[\mathrm{M}/\mathrm{H}]$ versus $[\alpha/\mathrm{M}]$, the Splash population largely overlaps with the thick disk sequence. These properties of the Splash stars further support the scenario proposed in the previous literature, which suggests that Splash stars formed in the Milky Way's high-$\alpha$ proto-disk and were subsequently heated to high-eccentricity orbits by a massive ancient accretion event \citep{Bonaca_2017, Haywood_2018, Matteo_2019, Amarante_2020, Belokurov_2020, Bonaca_2020}.

As shown in Figure~\ref{fig:GES1} and ~\ref{fig:Splash1}, our Splash population exhibits higher angular momentum and prograde rotational velocity, as well as lower orbital eccentricity, compared to the GES population. We initially distinguished Splash from GES based on these kinematic characteristics. However, we note that \citet{Tang_2024} found a Splash population within the GES stars identified in the IoM space. Additionally, Previous studies have suggested that Splash stars are generally high-$\alpha$ and metal-rich, consistent with the thick disk \citep{Belokurov_2020, Naidu_2020, Bonaca_2020, Deepak_2024}. Meanwhile, \citet{Zhao_2021}, based on the hydrodynamical simulation results of \citet{Amarante_2020}, suggested that both low-$\alpha$ and high-$\alpha$ MRSK stars coexist within GES. Therefore, we speculate that a small number of both low-$\alpha$ and high-$\alpha$ GES stars may be present in our Splash sample. This is also supported by the presence of a few low-$\alpha$ stars ($[\alpha/\mathrm{M}] < 0.2$\,dex) in the $[\mathrm{M}/\mathrm{H}]$ versus $[\alpha/\mathrm{M}]$ plane, as shown in Figure~\ref{fig:Splash1}.

The kinematic overlap between Splash and GES also implies the presence of dynamical interaction between them. Based on the above information, we suggest that the high-eccentricity, high-$\alpha$ stars may include both stars originating from the GES and Splash stars that were born in the Milky Way's high-$\alpha$ proto-disk and later heated onto eccentric orbits due to the GES merger event.

\subsection{The High-\texorpdfstring{$\alpha$}{alpha} Disk}
The high-$\alpha$ disk is a chemically distinct stellar population marked by elevated $\alpha$-element abundances \citep{Edvardsson_1993, Fuhrmann_1998, Chen_2000, Bensby_2003, Adibekyan_2012, Recio-Blanco_2014, Hayden_2015, Carollo_2019, Nissen_2020}. \citet{Bonaca_2020} selected 4,631 high-$\alpha$ stars with low orbital eccentricities ($e < 0.5$) from 11,000 main-sequence turn-off stars in the H3 Survey, and defined them as the “high-$\alpha$ disk”. Based on analyses of their orbital properties, ages, and elemental abundances, \citet{Bonaca_2020} proposed that this stellar population originated in the Milky Way’s high-$\alpha$ proto-disk. Subsequent studies have further identified low-eccentricity, $\alpha$-enhanced stellar populations as part of the high-$\alpha$ disk component, suggesting that these stars are typically old and formed during the early stages of the Galaxy's evolution \citep{Naidu_2020, Xiang_2022, Tang_2024, Xiang_2025}.

\begin{figure*}[!ht]
\centering
\includegraphics[width=0.7\textwidth]{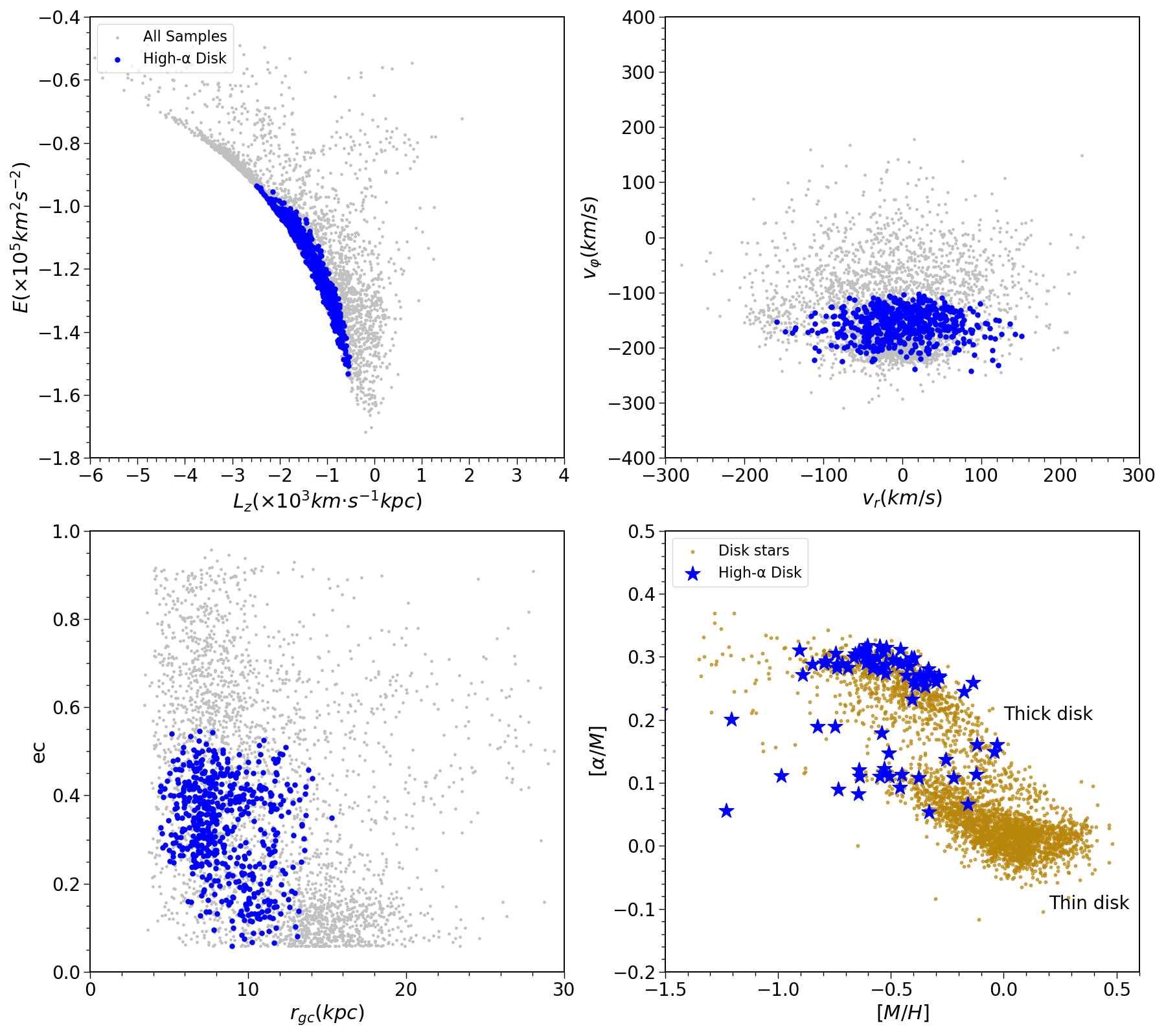}
\caption{\small Distribution of high-$\alpha$ disk stars in the $L_{\mathrm{z}}$-$E$, $V_{r}$-$V_{\phi}$, $r_{\mathrm{gc}}$-$ec$, and $[\mathrm{M}/\mathrm{H}]$-$[\alpha/\mathrm{M}]$ planes. The blue points represent high-$\alpha$ disk members, while the silver points represent all M giants. In the bottom-right panel, the blue stars represent 83 high-$\alpha$ disk stars, while the yellow points represent the disk stars from \citet{Yang_2019b}. All $[\mathrm{M}/\mathrm{H}]$ and $[\alpha/\mathrm{M}]$ measurements are taken from APOGEE DR17.
}\label{fig:High-α Disk1}
\end{figure*}

We identify 17 groups associated with the high-$\alpha$ disk, comprising a total of 557 stars. We cross-matched them with APOGEE DR17 and obtained 83 stars with reliable $[\mathrm{M}/\mathrm{H}]$ and $[\alpha/\mathrm{M}]$ measurements. As illustrated in Figure~\ref{fig:High-α Disk1}, the high-$\alpha$ disk spans from prograde circular orbits (\( L_z \sim -2.5 \times 10^3 \, \text{km} \cdot \text{s}^{-1} \cdot \text{kpc} \)) to elliptical orbits (\( L_z \sim -0.5 \times 10^3 \, \text{km} \cdot \text{s}^{-1} \cdot \text{kpc} \)) in the $L_z$-$E$ plane. The azimuthal velocity \(V_\phi\) and orbital eccentricity \(ec\) exhibit broad distributions, with \(V_\phi\) ranging from \(-100 \, \text{km} \, \text{s}^{-1}\) to \(-200 \, \text{km} \, \text{s}^{-1}\), and \(ec\) from 0.1 to 0.5, respectively.

Additionally, in the \([\mathrm{M}/\mathrm{H}]\)–\([\alpha/\mathrm{M}]\) plane, high-\(\alpha\) disk predominantly align with the thick disk sequence. The high rotational velocities, low orbital eccentricities, and thick disk-like chemical abundances collectively indicate that this stellar population formed in situ within the Milky Way’s proto-disk. This result further confirms the scenario proposed by \citet{Bonaca_2020}, which suggests that stars in the Milky Way initially formed in the high-$\alpha$ proto-disk. A subsequent massive ancient accretion event dynamically heated a fraction of these high-$\alpha$ stars, placing them on high-eccentricity orbits (now identified as the Splash population), while the remaining stars retained disk-like orbits, forming the so-called high-$\alpha$ disk.

Since the clustering analysis primarily relies on orbital parameters, some low-$\alpha$ disk stars with orbital features highly similar to those of the high-$\alpha$ disk population are misclassified into the latter (as shown in the lower right panel of Figure~\ref{fig:High-α Disk1}). These two populations can be further distinguished by incorporating more $\alpha$-abundance data.

\subsection{Unclassified Groups}
In our results, we also identify two groups which cannot be clearly associated with previously known substructures. We designate them as Group 1 and Group 2, containing 15 and 13 stars, respectively. We cross-matched the two groups with APOGEE DR17, and obtained two stars of Group 2 with reliable $[\mathrm{M}/\mathrm{H}]$ and $[\alpha/\mathrm{M}]$ measurements.

\begin{figure*}[!ht]
\centering
\includegraphics[width=0.7\textwidth]{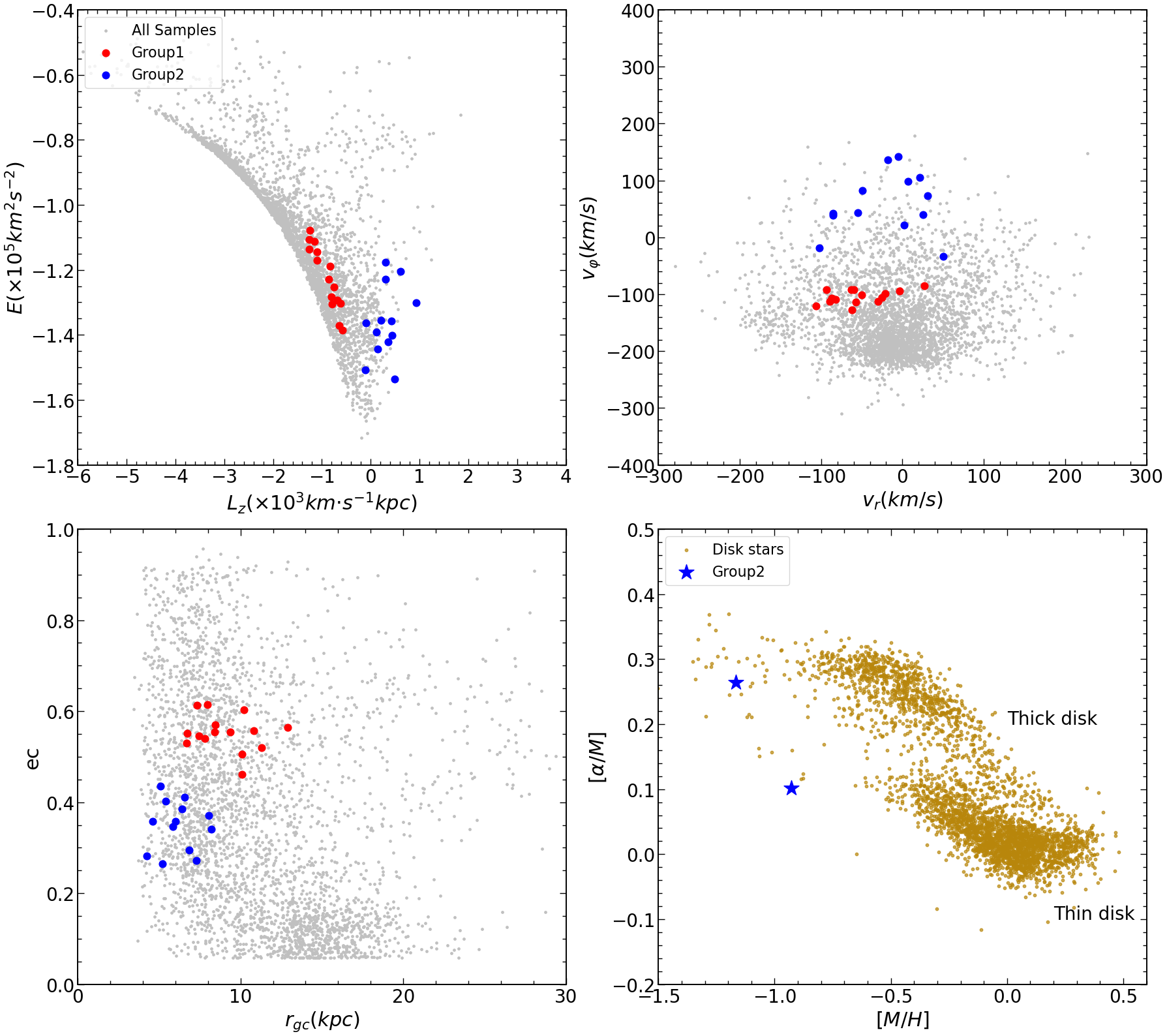}
\caption{\small The distribution of the unclassified groups in the chemodynamical space. The red and blue points in the figure represent Group 1 and Group 2, respectively. The silver points represent all M giants. In the bottom-right panel, the blue stars represent stars from Group 2, while the yellow points represent the disk stars from \citet{Yang_2019b}. All $[\mathrm{M}/\mathrm{H}]$ and $[\alpha/\mathrm{M}]$ measurements are taken from APOGEE DR17.
}\label{fig:new1}
\end{figure*}

Figure~\ref{fig:new1} illustrates the distribution of these groups in chemodynamical space. For Group 1, the mean z-component of angular momentum is $\langle L_z \rangle = -0.912 \times 10^3$ kpc km s$^{-1}$ and the mean energy is $\langle E \rangle = -1.22 \times 10^5$ km$^2$ s$^{-2}$. In contrast, Group 2 has a mean z-component of angular momentum $\langle L_z \rangle = 0.318 \times 10^3$ kpc km s$^{-1}$, and a mean energy $\langle E \rangle = -1.36 \times 10^5$ km$^2$ s$^{-2}$. In the $V_r$-$V_\phi$ plane, Group 1 is centered around $V_\phi = -100 \text{ km}~\text{s}^{-1}$, while Group 2 exhibits a more dispersed distribution within $0 < V_\phi < 200 \text{ km}~\text{s}^{-1}$, indicating slight retrograde motion. Additionally, Group 1 has a higher mean orbital eccentricity ($\langle ec \rangle = 0.55$) compared to Group 2 ($\langle ec \rangle = 0.35$).

In the kinematic space, Group 1 exhibits energy-angular momentum distributions and kinematic patterns similar to those of the Splash population. However, it has slightly lower orbital eccentricities and larger galactocentric distances ($r_{gc}$). Due to the lack of chemical abundance data, we cannot definitively determine the origin of Group 1. Nevertheless, its kinematic and dynamical properties suggest a potential origin in the Galactic disk.

Given its slight retrograde motion, we hypothesize that Group 2 is related to the low-energy retrograde substructure Thamnos \citep{Koppelman_2019b, Naidu_2020}, as it shares a relatively low energy $\langle E \rangle = -1.36 \times 10^5$ km$^2$ s$^{-2}$ and eccentricity $\langle ec \rangle = 0.35$. However, as shown in the lower-right panel of Figure~\ref{fig:new1}, two stars in Group 2 with metallicity measurements from APOGEE DR17 are more metal-rich than the classical Thamnos population, which typically has $[\mathrm{Fe}/\mathrm{H}] < -1.6$ dex \citep{Koppelman_2019b, Naidu_2020}. Nevertheless, we do not rule out the possibility that Group 2 represents the metal-rich tail of the Thamnos substructure.

\section{Discussion and Conclusion}\label{sect:Discussion and Conclusion}

In this study, we investigate Galactic substructures using a catalog of M giant stars from LAMOST DR9 low-resolution spectroscopic data (\citet{Li_2023}). By combining radial velocities from \citet{Li_2023}, distances from \citet{Qiu_2023}, and proper motions from Gaia DR3, we derive the full 6D phase-space information of these stars, including their spatial positions and velocity components. All M giants used for identifying substructure are available in Table~\ref{Table 1}.

After removing M giant stars with \(|Z_{gc}| < 2\) kpc, we apply the FoF clustering algorithm to identify stellar groups with similar orbital properties in the IoM space. This process yields 47 groups containing a total of 1,597 M giant stars. By comparing them with known substructures, we find that these groups primarily correspond to the Sgr stream (182 M giants), GASS (594 M giants), GES (115 M giants), Splash (121 M giants), and the high-$\alpha$ disk (557 M giants). Additionally, two groups cannot be clearly associated with previously known substructures. All orbital parameters of identified groups are listed in Table~\ref{Table 3}.

The Sgr stream in our sample consists of three components: the Sgr leading arm, the Sgr trailing arm, and Sgr debris. Their velocity and distance distributions are consistent with previous studies. Our analysis reveals that the traditionally defined leading stream consists of two kinematically distinct components, suggesting a more complex orbital structure than previously thought. Further investigation incorporating different types of Sgr member stars is required to better understand this complexity. The chemical abundance pattern of our Sgr member stars, characterized by low \([\alpha/\mathrm{M}]\) at high metallicity, aligns with previous findings and supports the chemical evolution trends commonly observed in Milky Way dwarf galaxies\citep{Venn_2012, Lemasle_2014, Yang_2019b}.

The GASS in our sample exhibits nearly circular prograde orbits spanning galactocentric distances from 10 to 20 kpc, consistent with the kinematic properties reported by \citep{Li_2021}. Chemically, the GASS stars have similar $\alpha$-element abundances to the thin disk but are comparatively more metal-poor, consistent with the continuation of the metal-rich thin disk population into the outer halo \citep{Haywood_2016, Hayes_2018, Li_2021, Zhang_2022}. These characteristics suggest that GASS originated from the disk and indicate a distinct formation and evolutionary history within the Milky Way's outer disk.

Analysis of the GES member stars reveals kinematic and chemical properties consistent with previous studies, including a bar-like distribution in $(L_z, E)$ space and a sausage-shaped structure in $(V_r, V_\phi)$ space with high eccentricities. The presence of metal-rich M giants exhibiting a wide range of $[\alpha/\mathrm{M}]$ supports the GES origin scenario proposed by \citet{Zhao_2021}, in which both the ancient GES merger event brought a gas-rich cloud into the Milky Way at an early epoch, and produced clumps that developed this bimodal chemistry for MRSK stars in later evolution. Furthermore, the spatial and dynamical similarities between our GES members and the HAC suggest a possible common origin, in agreement with the conclusions of \citet{Wang_2022}.

We identify the Splash component, characterized by their high orbital eccentricities and thick disk-like chemical abundance patterns. Our results demonstrate a clear kinematic distinction from thick disk stars and a similar chemical sequence, confirming the efficiency of our FoF clustering method in recovering Splash members in the IoM space. These findings support the scenario that Splash stars originated in the Milky Way’s high-$\alpha$ proto-disk and were dynamically heated by a massive early accretion event \citep{Bonaca_2017, Haywood_2018, Matteo_2019, Amarante_2020, Belokurov_2020, Bonaca_2020}. Additionally, the presence of a few low-$\alpha$ stars suggests a possible contamination from the low-$\alpha$, metal-rich tail of the GES galaxy \citep{Zhao_2021}, highlighting the complex interplay between in-situ and accreted populations. We also identify the high-$\alpha$ disk population, whose high rotational velocities, low orbital eccentricities, and thick disk-like chemical abundance patterns indicate an in situ origin from the Milky Way’s proto-disk.

In addition, we identify two groups which cannot be clearly associated with previously known substructures. Group~1 shows kinematic properties similar to the Splash but with lower orbital eccentricities and larger galactocentric distances, suggesting a possible origin in the Galactic disk. Group~2 shows kinematic consistency with the low-energy retrograde substructure Thamnos \citep{Koppelman_2019b, Naidu_2020}, though its higher metallicity suggests it may represent a more metal-rich extension of Thamnos.

Based on the kinematic and chemical properties of the metal-rich M giant stars successfully identified in this study as members of the GES, Splash, and high-$\alpha$ disk components, we confirm the evolutionary scenario of the early Milky Way proposed by previous studies, in which stars in the Milky Way initially formed in the high-$\alpha$ proto-disk. A subsequent massive ancient accretion event dynamically heated a fraction of these high-$\alpha$ stars, placing them on high-eccentricity orbits (now identified as the Splash population), while the remaining stars retained disk-like orbits, forming the so-called high-$\alpha$ disk \citep{Helmi_2018, Belokurov_2020, Bonaca_2020, Zhao_2021, Chandra_2024}. In the future, more precise measurements of their chemical abundances (e.g., $\mathrm{[\alpha/M]}$ and $[\mathrm{Fe}/\mathrm{H}]$) using high-resolution spectroscopic data, along with age estimates, will be crucial in robustly verifying the origin of these M giants and their role in the GES merger event. This discovery provides new insights into the star formation history of the GES progenitor galaxy and its contribution to the evolution of the Milky Way.

\begin{table}[!ht]
    \centering
    \caption{\small Orbital parameters of all Galactic substructures we identified. The first column indicates the stellar substructure affiliation, and the second column represents the LAMOST observation ID.
}
    \resizebox{0.96\textwidth}{!}{ 
    \begin{tabular}{cccccccccccc}
    \hline
        Substructure & LAMOST & a & ec & $l_{\text{or}}$ & $b_{\text{or}}$ & $l_{\text{apo}}$ & E & L & $L_{\text{x}}$ & $L_{\text{y}}$ & $L_{\text{z}}$  \\ \hline
        Sgr stream & 270210167 & 11.78 & 0.25 & 227.59 & -54.34 & 47.52 & -95226.95 & 2341 & -919.74 & -1007.26 & -1901.75  \\ 
        Sgr stream & 563808192 & 27.84 & 0.33 & 279.15 & -14.23 & 100.35 & -65980.43 & 4770.51 & 735.31 & -4564.55 & -1169.49  \\ 
        Sgr stream & 814912143 & 30.28 & 0.36 & 283.34 & -13.32 & 99.3 & -63337.47 & 5034.6 & 1130.96 & -4767.55 & -1158.31  \\ 
        Sgr stream & 660004143 & 14.77 & 0.37 & 224.66 & -30.11 & 28.61 & -86717.55 & 2659.09 & -1614.48 & -1616.21 & -1334.45  \\ 
        Sgr stream & 531309081 & 31.46 & 0.29 & 285.95 & -17.66 & 108.47 & -62335.35 & 5393.26 & 1409.21 & -4939.6 & -1637.71  \\ 
        Sgr stream & 422602068 & 14.16 & 0.34 & 221.21 & -29.62 & 43.01 & -88327.87 & 2624.12 & -1699.25 & -1503.97 & -1298.09  \\ 
        Sgr stream & 700501081 & 32.02 & 0.31 & 279.85 & -13.93 & 95.23 & -61789.63 & 5445.59 & 903.74 & -5207.6 & -1313.01  \\ 
        Sgr stream & 425310085 & 26.88 & 0.43 & 269.53 & -9.74 & 86.26 & -66773.76 & 4298.02 & -34.1 & -4234.39 & -728.34  \\ 
        Sgr stream & 502013024 & 34.04 & 0.31 & 283.97 & 0.22 & 83.47 & -59869.61 & 5790.92 & 1401 & -5617.93 & 22.64  \\ 
        Sgr stream & 132105073 & 24.01 & 0.4 & 269.08 & -18.68 & 80.35 & -70471.15 & 3974.71 & -60.54 & -3761.52 & -1280.21  \\ \hline
    \end{tabular}
    }
    \label{Table 3}
    \begin{flushleft}
    (This table is available in its entirety in machine-readable form.)
    \end{flushleft}
\end{table}

\begin{acknowledgements}

We thank the referee for providing a number of helpful suggestions that further refined our work. This work is supported by the National Key Research and Development Program of China No. 2024YFA1611902, the National Natural Science Foundation of China (NSFC) under grants Nos. 12273027 and 12588202, the Innovation Team Funds of China West Normal University under grant No. KCXTD2022-6, CAS Project for Young Scientists in Basic Research under grants Nos. YSBR-062 and YSBR-092, and the Strategic Priority Research Program of Chinese Academy of Sciences under grant No. XDB1160102. We acknowledge the science research grant from the China Manned Space Project with No. CMS-CSST-2025-A11.

Guo Shou Jing Telescope (the Large Sky Area Multi-Object Fiber Spectroscopic Telescope LAMOST) is a National Major Scientific Project built by the Chinese Academy of Sciences. Funding for the project has been provided by the National Development and Reform Commission. LAMOST is operated and managed by the National Astronomical Observatories, Chinese Academy of Sciences.

This work has made use of data from the European Space Agency (ESA) mission Gaia (\url{https://www.cosmos.esa.int/gaia}), processed by the Gaia Data Processing and Analysis Consortium (DPAC, \url{https://www.cosmos.esa.int/web/gaia/dpac/consortium}). 
Funding for the DPAC has been provided by national institutions, in particular the institutions participating in the Gaia Multilateral Agreement.
\end{acknowledgements}

\appendix
\section{Determination of the Linking Length}\label{APP:A}
Considering the normalized weights of the input sample and the characteristic linking lengths of different substructures, the linking length for each group is determined individually\citep{Wang_2022}. As the linking length increases, the number of group members grows slowly at first, but a sudden jump occurs at a certain linking length. This jump likely indicates the merging of multiple groups, leading to the combination of several substructures\citep{Zhang_2024}. To avoid such cases, we choose the linking length just before the jump by checking the orbital parameter distributions of the group before and after the jump. 

\renewcommand{\thefigure}{A\arabic{figure}}
\setcounter{figure}{0}

\begin{figure*}[!ht]
\centering
\includegraphics[width=0.65\textwidth]{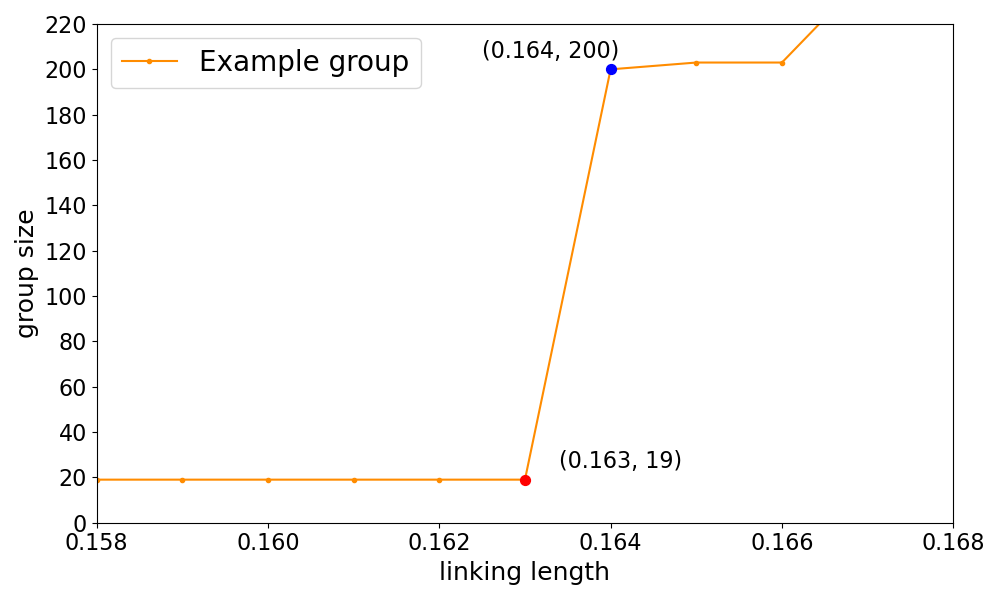}
\caption{\small The variation in the number of members of the example group with increasing linking length.
}\label{fig:figureA1}
\end{figure*}

\begin{figure*}[!ht]
\centering
\includegraphics[width=0.8\textwidth]{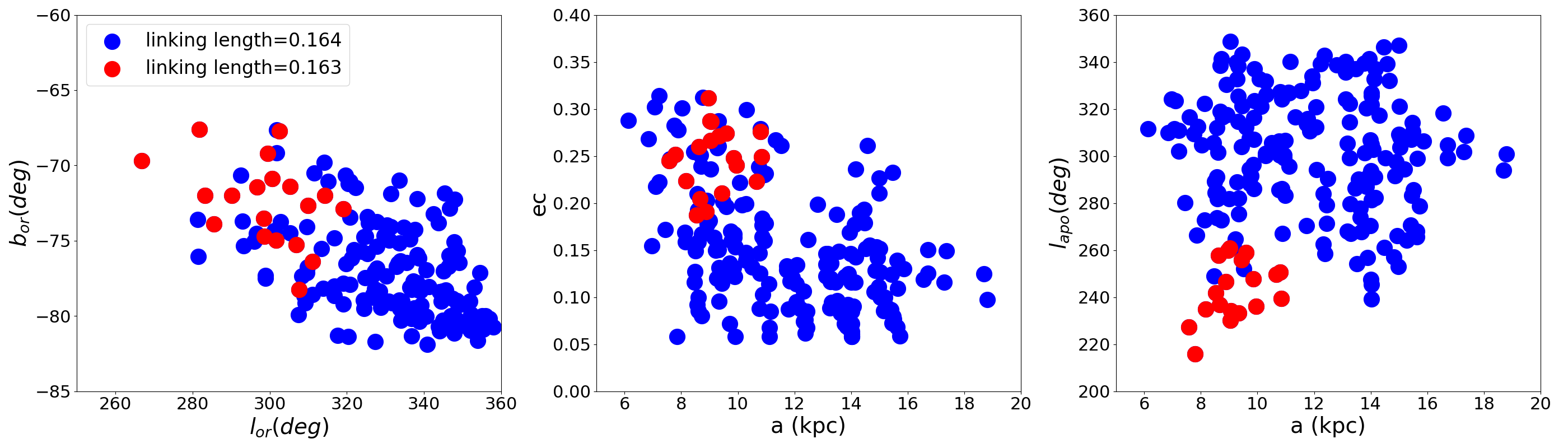}
\caption{\small The orbital parameter distribution of the example group members. The red points represent the example group members at a linking length of 0.163, while the blue points represent the example group members at a linking length of 0.164.
}\label{fig:figureA2}
\end{figure*}

Figure~\ref{fig:figureA1} illustrates how the number of members in an example group changes with increasing linking length. It shows a significant jump in the number of members when the linking length increases from 0.163 to 0.164. Figure~\ref{fig:figureA2} displays the orbital parameter distributions of the group members before and after the jump. It can be seen that at linking length = 0.164, the group becomes unreliable, particularly as the distribution of \( l_{\rm apo} \) suggests the presence of multiple substructures. Therefore, we retain this group at linking length = 0.163.

\section*{ORCID iDs}
\begin{flushleft}
\author{Longfei Ding\, 
\orcidlink{0009-0007-6842-8117}} \href{https://orcid.org/0009-0007-6842-8117}{https://orcid.org/0009-0007-6842-8117}

\author{Jing Li\, 
\orcidlink{0000-0002-4953-1545}} \href{https://orcid.org/0000-0002-4953-1545}{https://orcid.org/0000-0002-4953-1545}

\author{Xiang-Xiang Xue\, 
\orcidlink{0000-0002-0642-5689}} \href{https://orcid.org/0000-0002-0642-5689}{https://orcid.org/0000-0002-0642-5689}

\author{Hao Tian\, 
\orcidlink{0000-0003-3347-7596}} \href{https://orcid.org/0000-0003-3347-7596}{https://orcid.org/0000-0003-3347-7596}

\author{Zhengzhou Yan\, 
\orcidlink{0000-0003-3571-6060}} \href{https://orcid.org/0000-0003-3571-6060}{https://orcid.org/0000-0003-3571-6060}

\author{Gang Zhao\, 
\orcidlink{0000-0002-8980-945X}} \href{https://orcid.org/0000-0002-8980-945X}{https://orcid.org/0000-0002-8980-945X}
\end{flushleft}

\bibliographystyle{aasjournal}
\bibliography{main}

\label{lastpage}

\end{document}